\documentstyle[seceq,epsf]{ptptex}
\title{%        %You can use \\ for explicit line-break
Microscopic Study of Nuclear ``Pasta''
by\\ Quantum Molecular Dynamics
}
\author{%       %Use \sc for the family name
Gentaro {\sc Watanabe}$^{a,b}$,
Katsuhiko {\sc Sato}$^{a,c}$,
Kenji {\sc Yasuoka}$^{d}$ and
Toshikazu {\sc Ebisuzaki}$^{b}$
}

\inst{%         %Affiliation, neglected when [addenda] or [errata]
$^{a}$Department of Physics, University of Tokyo,
Tokyo 113-0033 %, Japan
\\
$^{b}$Division of Computational Science, RIKEN,
Saitama 351-0198 %, Japan
\\
$^{c}$Research Center for the Early Universe, 
University of Tokyo,
Tokyo 113-0033 %, Japan
\\
$^{d}$Department of Mechanical Engineering, Keio University,
Yokohama 223-8522 %, Japan
}

\recdate{%      %Editorial Office will fill in this.
}

\abst{%       %this abstract is neglected when [addenda] or [errata]
Structure of cold dense stellar matter
at subnuclear densities is investigated
with quantum molecular dynamics (QMD).
We succeeded in showing that the phases with slab-like and
rod-like nuclei etc. can be formed dynamically
without any assumptions on the nuclear shape.
Our result suggests the existence of these kinds of phases
in neutron star crusts.
}

\begin{document}

\maketitle

\section{Introduction}
%Start your paper from here.
Collapse driven supernovae (SNe)
and the following formation of neutron stars (NSs)
are the most dramatic processes during the stellar evolution.
These objects provide not only astrophysically significant phenomena
but also interesting material phases inside them;
both are strongly connected with each other.
At subnuclear densities
in NS crusts and SN cores, nuclei are expected to have exotic structures
such as rod-like and slab-like shape etc.,
which are called nuclear ``pasta''.\cite{rf:review}
This prediction has been obtained
by previous studies \cite{rf:rpw,rf:hashimoto}
based on free energy calculations assuming specific nuclear shapes.
These works clarify that the nuclear shape is
determined by a subtle balance
between nuclear surface and Coulomb energies.

However, the formation process of the ``pasta'' phases has not been discussed
except for some limited cases which are based on perturbative approaches.
\cite{rf:review,rf:iida}
It is important to adopt microscopic and dynamical approach which allows
arbitrary nuclear structures in order to understand the formation
of the non-spherical nuclei.
We have started studying dense matter
by the QMD\cite{rf:qmd} simulation
whose final goal is to obtain a realistic picture
about NS crusts and SN cores.
As the first step of our quest,
we will answer the following question in this study:
``Can the 'pasta' phases be formed dynamically?''

\section{Simulations and results}
In the formation process of the ``pasta'' phases,
density fluctuations would be crucial.
Methods of treating fermion many body systems
based on molecular dynamics
are suitable for incorporating fluctuations of particle distribution.
\begin{figure}
  \epsfysize=7 pc
  \centerline{\epsfbox{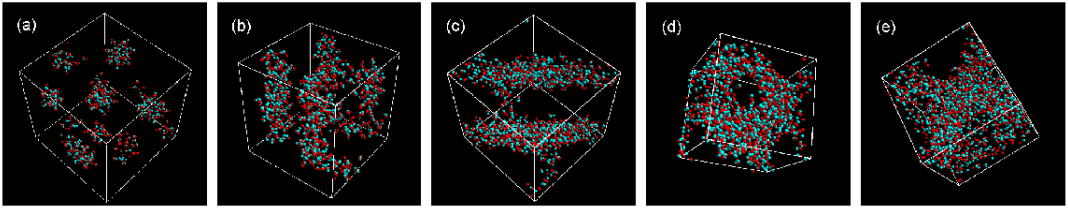}}
  \caption{Nucleon distibutions of cold symmetric nuclear matter
    at various densities; (a) sphere phase, $0.1 \rho_{0}$;
    (b) cylinder phase, $0.25 \rho_{0}$; (c) slab phase, $0.4 \rho_{0}$;
    (d) cylindrical hole phase, $0.5 \rho_{0}$ and
    (e) spherical hole phase, $0.6 \rho_{0}$.
    The red particles represent protons and green ones represent neutrons.
    Temperatures are $O(0.1)$ MeV.}
  \label{fig:1}
\end{figure}
We choose the QMD among a lot of versions of MD for fermions
based on trade-off between the calculational amounts and accuracies.
It is noted that we watch macroscopic structure,
for which the exchange effect is not so important.
Therefore, it is justified to use the QMD
which is less elaborate in treating the exchange effect.

We have performed QMD simulations of (n,p,e) system
with proton fraction $x=0.3$ and 0.5
for various nucleon densities $\rho$.
We set a cubic box
which is imposed periodic boundary conditions.
Total nucleon numbers are 1372 and 2048.
The Coulomb interaction is calculated by the Ewald method
and relativistic electrons are treated as a uniform background.
For the nucleon interaction,
we use a model Hamiltonian developed by Maruyama et al.\cite{rf:maruyama}

We first prepare uniform hot nucleon gas
at $k_{B}T \sim 20$ MeV as an initial condition,
and then cool it down slowly for $O(10^{3}-10^{4})$ fm/c
until the temperature gets $O(0.1)$ MeV.
Note that any artificial fluctuations are not given
during the simulation.
Shown in Fig.\ 1 is the resultant nucleon distributions of
cold symmetric matter for $\rho = 0.25-0.6 \rho_{0}$
(\ $\rho_{0}$ : the normal nuclear density\ ).
We can see from Fig.\ 1 that the phases with rod-like and
slab-like nuclei, cylindrical and spherical bubbles,
in addition to the one with spherical nuclei, are reproduced.
%; the sequence of the nuclear shape is consistent with the results of
%the previous studies.
Also for $x=0.3$, these phases are reproduced
at about $0.05 \rho_{0}$ lower than the symmetric matter case.
%Also for $x=0.3$, these phases are reproduced
%with shifting to the lower density side about $0.05 \rho_{0}$
%from the case of the symmetric matter.

The density dependence of the nuclear shape is quite sensitive
and various intermediate phases are observed
in density regions between the densities
corresponding to the phases with the simple shapes
as shown in Fig.\ 1.
Whole phase diagrams of cold matter
and properties of the nuclear shape changes
will be discussed elsewhere.\cite{rf:gentaro}
It is concluded that our results suggest
the existence of the ``pasta'' phases in NS crusts
which cool down keeping the local thermal equilibrium
after the proto-NS is formed,
and their cooling time scale is much longer than that of our simulations.
%and their cooling time scale is macroscopic
%and is much longer than that of our simulations.

%\section*{Acknowledgements}

\end{document}